\documentclass[11pt,a4paper]{article}
\pdfoutput=1

\usepackage[numbers, square, comma, sort&compress]{natbib}
\usepackage{amssymb}
\usepackage{amsthm}
\usepackage{amstext}
\usepackage{amsmath}
\usepackage{amsfonts}
\usepackage{empheq}
\usepackage{verbatim}
\usepackage{geometry}
\usepackage{latexsym}
\usepackage{t1enc}
\usepackage{graphicx}
\usepackage[all]{xy}
\usepackage{makeidx}
\usepackage{slashed}
\usepackage{multicol}
\usepackage{alltt}

\allowdisplaybreaks

\newcommand{\as}{\alpha_s}
\newcommand{\be}{\begin{equation}}
\newcommand{\ee}{\end{equation}}
\newcommand{\bea}{\begin{eqnarray}}
\newcommand{\eea}{\end{eqnarray}}
\newcommand{\eps}{\epsilon}

\newcommand{\e}{\epsilon}

\newcommand{\secn}[1]{Section~\ref{#1}}

\def\beq{\begin{equation}}
\def\eeq{\end{equation}}
\def\beqa{\begin{eqnarray}}
\def\eeqa{\end{eqnarray}}
\def\eq#1{Eq.~(\ref{#1})}

\def\slash#1{#1 \hskip-0.45em /}

\def\spa#1.#2{\left\langle#1\,#2\right\rangle}
\def\spb#1.#2{\left[#1\,#2\right]}
\def\spash#1.#2{\spa{\smash{#1}}.{\smash{#2}}}
\def\spbsh#1.#2{\spb{\smash{#1}}.{\smash{#2}}}
\def\sand#1.#2.#3{%
  \left\langle\smash{#1^{-}}{\vphantom1}\right|{#2}%
  \left|\smash{#3^{-}}{\vphantom1}\right\rangle}
\def\sandp#1.#2.#3{%
  \left\langle\smash{#1^{-}}{\vphantom1}\right|{#2}%
  \left|\smash{#3^{+}}{\vphantom1}\right\rangle}
\def\sandpp#1.#2.#3{% 
  \left\langle\smash{#1^{+}}{\vphantom1}\right|{#2}%
  \left|\smash{#3^{+}}{\vphantom1}\right\rangle}
\def\sandpm#1.#2.#3{%
  \left\langle\smash{#1^{+}}{\vphantom1}\right|{#2}%
  \left|\smash{#3^{-}}{\vphantom1}\right\rangle}
\def\sandmp#1.#2.#3{%
  \left\langle\smash{#1^{-}}{\vphantom1}\right|{#2}%
  \left|\smash{#3^{+}}{\vphantom1}\right\rangle}

\def\ssand#1.#2.#3{%
  \left\langle\smash{#1}{\vphantom1}\right|{#2}%
  \left|\smash{#3}{\vphantom1}\right]}
\def\ssandp#1.#2.#3{%
  \left\langle\smash{#1}{\vphantom1}\right|{#2}%
  \left|\smash{#3}{\vphantom1}\right\rangle}
\def\ssandpp#1.#2.#3{%
  \left\langle\smash{#1}{\vphantom1}\right|{#2}%
  \left|\smash{#3}{\vphantom1}\right\rangle}

\def\proj{\flat}
\def\projdot#1.#2{k_{#1}^\proj\cdot k_{#2}^\proj}
\def\sandff#1.#2.#3{%
  \left\langle\smash{#1^{\proj,-}}{\vphantom1}\right|{#2}%
  \left|\smash{#3^{\proj,-}}{\vphantom1}\right\rangle}
\def\sandnf#1.#2.#3{%
  \left\langle\smash{#1^{-}}{\vphantom1}\right|{#2}%
  \left|\smash{#3^{\proj,-}}{\vphantom1}\right\rangle}
\def\sandfn#1.#2.#3{%
  \left\langle\smash{#1^{\proj,-}}{\vphantom1}\right|{#2}%
  \left|\smash{#3^{-}}{\vphantom1}\right\rangle}

\def\spa#1.#2{\left\langle#1\,#2\right\rangle}
\def\spb#1.#2{\left[#1\,#2\right]}

\numberwithin{equation}{section}

\begin{document}

%%%%%%%%%%%%%%%%%%%%%%%%%%%%%%%%%%%%%%%%%%%%%

\begin{titlepage}

\hbox{TTK-16-40}
\hbox{NIKHEF/2016-043}
\hbox{NSF-KITP-15-145}
\hbox{Edinburgh 2016/16}
\hbox{QMUL-PH-16-17}

\vskip 25mm

\begin{center}
\Large{\sc{Non-abelian factorisation for next-to-leading-power threshold logarithms}}
\end{center}

\vskip 8mm

\begin{center}

D.~Bonocore$^{1,2}$, E.~Laenen$^{1,3,4,5}$, L.~Magnea$^6$, 
L.~Vernazza$^7$, C.~D.~White$^8$ \\ [6mm]

\vspace{6mm}

\textit{$^1$Nikhef, Science Park 105, NL--1098 XG Amsterdam, The Netherlands} \\ 
\vspace{1mm}

\textit{$^2$Institute for Theoretical Particle Physics and Cosmology, RWTH Aachen University,
Sommerfeldstr. 16, 52074 Aachen, Germany}\\
\vspace{1mm}

\textit{$^3$ITFA, University of Amsterdam, Science Park 904, Amsterdam, 
The Netherlands} \\
\vspace{1mm}

\textit{$^4$ITF, Utrecht University, Leuvenlaan 4, Utrecht, The Netherlands} \\
\vspace{1mm} 

\textit{$^5$Kavli Institute for Theoretical Physics, University of California,\\ 
Santa Barbara, CA 93106-4030} \\
\vspace{1mm}

\textit{$^6$Dipartimento di Fisica, Universit\`a di Torino and INFN, Sezione di Torino \\
Via P. Giuria 1, I-10125 Torino, Italy} \\
\vspace{1mm}

\textit{$^7$Higgs Centre for Theoretical Physics, School of Physics and Astronomy, 
The University of Edinburgh, Edinburgh EH9 3JZ, Scotland, UK}\\
\vspace{1mm}

\textit{$^8$Centre for Research in String Theory, School of Physics and Astronomy, 
Queen Mary University of London, 327 Mile End Road, London E1 4NS, UK} \\
\vspace{1mm}

\end{center}

\vspace{5mm}

\begin{abstract}

\noindent
Soft and collinear radiation is responsible for large corrections to many 
hadronic cross sections, near thresholds for the production of heavy final 
states. There is much interest in extending our understanding of this 
radiation to next-to-leading power (NLP) in the threshold expansion. 
In this paper, we generalise a previously proposed all-order NLP factorisation 
formula to include non-abelian corrections. We define a non-abelian radiative 
jet function, organising collinear enhancements at NLP, and compute it for
quark jets at one loop. We discuss in detail the issue of double counting 
between soft and collinear regions. Finally, we verify our prescription by 
reproducing all NLP logarithms in Drell-Yan production up to NNLO, 
including those associated with double real emission. Our results constitute 
an important step in the development of a fully general resummation 
formalism for NLP threshold effects.

\end{abstract}

\end{titlepage}

%%%%%%%%%%%%%%%%%%%%%%%%%%%%%%%%%%%%%%%%%%%%%

\section{Introduction}
\label{sec:introduction}

The properties of QCD radiation near partonic thresholds for the production of
heavy final states have a significant impact on a wide range of phenomenologically 
relevant collider observables. Typically, if $\xi$ is a dimensionless kinematic variable 
vanishing at the threshold, differential QCD cross-sections will contain terms of the 
form
\be
  \frac{d \sigma}{d \xi} \, = \, \sum_{n = 0}^{\infty} \left( \frac{\alpha_s}{\pi} \right)^n 
  \sum_{m = 0}^{2 n - 1} \left[ \, c_{n m}^{(-1)} \left( \frac{\log^m \xi}{\xi} \right)_+ + 
  \, c_{nm}^{(\delta)} \, \delta(\xi) + \, c_{nm}^{(0)} \, \log^m \xi + \ldots \, \right] \, ,
\label{thresholddef}
\ee
where the ellipsis denotes terms suppressed by further powers of $\xi$. The first 
set of terms, at leading power in $\xi$, originates from the singularities associated
with soft and collinear gluon emission. These singularities are universal and factorising,
which leads to the possibility of resumming the resulting logarithms to all orders in 
perturbation theory. The formalism to perform this resummation is well-known, and 
it has been extensively applied to a plethora of collider observables (see, for example,
\cite{Sterman:1986aj,Catani:1989ne,Korchemsky:1993uz,Korchemsky:1993xv,
Forte:2002ni,Contopanagos:1997nh,Banfi:2004yd,Becher:2006nr,Luisoni:2015xha}). 
The second set of terms in \eq{thresholddef}, which are localised at threshold, 
originates mostly from singular virtual corrections to the production amplitude. 
These terms can also be organised to all orders for processes which are electroweak 
at tree level~\cite{Parisi:1980xd,Eynck:2003fn,Ahrens:2008qu}, albeit with reduced
predictive power. The vast amount and increased precision of LHC data, together 
with the lack of any striking signature for new physics, make the third set of terms 
in \eq{thresholddef}, at next-to-leading power in the threshold parameter $\xi$, 
potentially relevant for precision Standard Model studies. Indeed, quite a large
body of work has already been devoted to this problem.

The fact that at least some NLP contributions can be understood to all
orders is well known, as a consequence of the LBKD
theorem~\cite{Low:1958sn,Burnett:1967km, DelDuca:1990gz}. Further
evidence for a non-trivial relation between LP and NLP logarithms came
from the analysis of DGLAP splitting functions in
Ref.~\cite{Dokshitzer:2005bf}.  Since then, and following early
studies in~\cite{Kramer:1996iq,Akhoury:1998gs}, several groups have
attempted to construct a systematic formalism for understanding NLP
logarithms, using a variety of methods, ranging from path integral
techniques~\cite{Laenen:2008gt}, to diagrammatic
approaches~\cite{Laenen:2010uz}, physical evolution
kernels~\cite{Soar:2009yh,
  Moch:2009hr,Moch:2009mu,Almasy:2010wn,deFlorian:2014vta,Presti:2014lqa},
effective field theories~\cite{Larkoski:2014bxa,Kolodrubetz:2016uim},
and other techniques~\cite{Laenen:2008ux,
  Grunberg:2009yi,Grunberg:2009vs}. Interestingly, the study of
next-to-soft contributions to scattering amplitudes in both gauge
theory and gravity from a more formal point of view, based on
asymptotic symmetries of the S matrix, has also received a great deal
of attention (see for
example~\cite{Strominger:2013jfa,Cachazo:2014fwa,
  Casali:2014xpa,Bern:2014oka,Larkoski:2014hta,White:2011yy,White:2014qia}).

Recently, building on the results of~\cite{DelDuca:1990gz}, in Ref.~\cite{Bonocore:2015esa}
we proposed a factorisation formula for the Drell-Yan scattering amplitude, valid at the
accuracy needed to generate NLP logarithms in the cross section. This formula
generalises the factorisation of soft and collinear divergences by including NLP effects,
and contains the same universal functions as the leading-power factorisation, together 
with a new universal {\it radiative jet function}, responsible for next-to-soft emission 
from a collinearly enhanced configuration. Ref.~\cite{Bonocore:2015esa} evaluated 
this quantity up to one-loop order for an external quark, and, using as a guideline the 
calculation performed with the method of regions~\cite{Beneke:1997zp,Pak:2010pt,
Jantzen:2011nz} in Ref.~\cite{Bonocore:2014wua}, succeeded in reproducing 
a set of NLP terms in the Drell-Yan cross section at NNLO, originally computed 
in~\cite{Matsuura:1988nd,Matsuura:1989sm}.

The factorisation formula proposed in Refs.~\cite{DelDuca:1990gz,Bonocore:2015esa} 
was, however, appropriate for an abelian theory, and could only reproduce abelian-like
QCD contributions, proportional  to the color factor $C_F^n$ at ${\cal O}(\alpha_s^n)$. 
The aim of this paper is to provide a fully non-abelian NLP factorisation formula, a 
generalisation from previous results which is non-trivial for a number of reasons.

First, it is necessary to include the emission of colour-correlated gluons from outside 
the hard interaction. These diagrams were called {\it next-to-eikonal webs} in 
Refs.~\cite{Laenen:2008gt,Laenen:2010uz}, where they were shown to be described 
by generalised Wilson line operators, obeying exponentiation properties similar to their 
leading-power counterparts. Second, the definition of the radiative jet function must 
be generalised to cope with a non-abelian operator insertion for the additional gluon. 
Third, one must address double counting of soft and collinear regions at NLP level,
a problem which occurs also at leading power, or in effective field theory
approaches~\cite{Larkoski:2014bxa,Kolodrubetz:2016uim}, but was easily circumvented in the abelian 
limit. 

The structure of our paper is as follows. In \secn{sec:factorisation}, we introduce
a new, non-abelian definition of the radiative jet function, in terms of a non-abelian
conserved current, and we use it to derive a complete factorisation formula for 
the Drell-Yan amplitude, with the required accuracy to reproduce all NLP effects.
In \secn{sec:jetemit}, we compute the new radiative jet function to one loop, which
is sufficient to generate all NLP logarithms in the cross section at NNLO. In
\secn{sec:DY}, we use these results to check that the known non-abelian terms 
in the Drell-Yan K-factor up to NNLO are indeed reproduced. In \secn{sec:tworeal}, 
we briefly describe how our methods can also reproduce double-real emission 
contributions at NNLO. In \secn{sec:conclude} we present our conclusion and 
outline future work towards an effective NLP resummation formalism.

%%%%%%%%%%%%%%%%%%%%%%%%%%%%%%%%%%%%%%%%%%%%%

\section{The non-abelian NLP factorisation formula}
\label{sec:factorisation}

%%%%%%%%%%%%%%%%%%%%%%%%%

\subsection{Leading power factorisation}
\label{sec:LP}

Anticipating the Drell-Yan application of \secn{sec:DY}, we consider a quark scattering 
amplitude involving two partons with momenta $p_1$ and $p_2$, which we write as
\be
  {\cal M} (p_1, p_2) \, = \, \bar{v}(p_2) \, {\cal A} (p_1, p_2) \, u(p_1) \, ,
\label{Mdef}
\ee
so that ${\cal A}$ has the external fermion wave functions removed. In massless QCD,
${\cal A}$ is affected by infrared and collinear divergences, which however factorise to 
all orders in the form~\cite{Dixon:2008gr}
\be 
  {\cal A} (p_1,p_2) \, = \, {\cal H} \left( p_j, n_j \right) \, {\cal S} \left( \beta_j \right) \,
  \prod_{i = 1}^2 \frac{ J(p_i, n_i) }{ {\cal J} (p_i, n_i)} \, .
\label{softcolfac}
\ee
In \eq{softcolfac}, $J(p_i, n_i)$ is a {\it jet function}, collecting collinear singularities 
associated with parton $i$: it depends on an auxiliary vector $n_i$, as described below;
this dependence cancels with the other factors in \eq{softcolfac}, so that the full 
scattering amplitude is independent of $n_i$, as expected. For a quark, the jet function
is given by~\footnote{Throughout, we leave time ordering implicit for brevity.}
\beq
  J( p, n) u(p) \, = \, \left\langle 0 \left| \Phi_n (\infty, 0)
  \psi(0) \right| p \right\rangle \, ,
\label{Jdef}
\eeq
where the fermion field $\psi(x)$ absorbs the external parton of momentum $p$, 
and $\Phi_n (\infty, 0)$ is a Wilson line in the direction of the auxiliary four vector 
$n_\mu$, guaranteeing gauge invariance, and defined according to
\be
  \Phi_v ( \lambda_2, \lambda_1) \, = \, {\cal P} \exp \left[ i
  g \int_{\lambda_1}^{\lambda_2} d \lambda\, v \cdot
  A(\lambda v) \right] \, .
\label{phidef}
\ee
The {\it soft function} ${\cal S} (\beta_i)$ collects soft divergences, and is a correlator 
of Wilson lines directed along the classical trajectories of the hard emitting particles: 
in fact, $\beta_i$ is a dimensionless vector proportional to the four-velocity of parton 
$i$ according to $p_i = Q \beta_i$, with $Q$ a hard scale. We define then
\beq
  {\cal S} \left( \beta_i \right) \, = \, \left\langle
  0 \left| \Phi_{\beta_2} (\infty,0) \Phi_{\beta_1} (0,
  -\infty) \right| 0 \right\rangle \, .
\label{Sdef}
\eeq
The final ingredient of \eq{softcolfac}, the {\it eikonal jet} function ${\cal J} (p_i, n_i)$,
is responsible for subtracting the double counting of soft-collinear configurations, 
which contribute to both the jets and the soft function. The eikonal jet is obtained 
by replacing the hard line of momentum $p^\mu$ in the partonic jet with a Wilson line 
with four-velocity $\beta^\mu$, yielding
\beq
  {\cal J}( \beta, n) \, = \, \left\langle 0 \left|
  \Phi_n (\infty, 0) \Phi_{\beta} (0, -\infty) \right| 0 \right\rangle \, .
\label{eikJdef}
\eeq
After factorising all singular (and universal) contributions, the matching to the exact 
amplitude order by order yields ${\cal H}$, an infrared-finite, but process-dependent 
{\it hard function}.

\eq{softcolfac} forms the starting point for describing radiation at leading power 
in the threshold expansion, where the soft function corresponds to dressing the 
hard function with virtual gluons of 4-momentum $k^\mu \rightarrow 0$, whose 
infrared singularities cancel those associated with real emissions. In what follows,
considering real radiation up to next-to-soft level, it will be convenient to generalise 
the soft function to include sub-leading powers of momentum in the propagators 
and emission vertices for virtual gluons. This can be done by defining a 
{\it next-to-soft function} as
\beq
  \widetilde{\cal S} \left( p_1, p_2 \right) \, = \, 
  \left. \left\langle 0 \left| F_{p_2} (\infty,0) F_{p_1} (0, -\infty) 
  \right| 0 \right\rangle\right|_{\rm NLP} \, , 
\label{Stildedef}
\eeq
where $F_p$ is a generalised Wilson line operator, constructed in Ref.~\cite{Laenen:2008gt}, 
which generates the required next-to-soft emission vertices. A general definition of such an
operator for generic trajectories can be given in a coordinate-space representation, and 
is presented in~\cite{Laenen:2008gt}. For straight semi-infinite trajectories one can easily
transform to a momentum-space representation, given by
\beqa
  F_p (0,\infty) & = & {\cal P} \exp \left[ \, g \! \int \frac{d^d k}{(2 \pi)^d} A_\mu(k)
  \left( - \frac{p^\mu}{p\cdot k} + \frac{k^\mu}{2 p \cdot k} - 
  k^2 \frac{p^\mu}{2 (p \cdot k)^2} - \frac{{\rm i} k_\nu \Sigma^{\nu\mu}}
  {p \cdot k} \right) \right. \nonumber \\
  && \,\, \, + \left. \int \frac{d^d k}{(2 \pi)^d} \int \frac{d^d l}{(2 \pi)^d}
  A_\mu(k) A_\nu(l) \left(\frac{\eta^{\mu \nu}}{2 p \cdot (k + l)} - 
  \frac{p^\nu l^\mu p \cdot k + p^\mu k^\nu p \cdot l}{2 (p \cdot l)(p \cdot k) 
  \left[p \cdot (k + l) \right]} \right.\right. \nonumber \\
  && \,\,\,  + \left. \left. \frac{(k \cdot l) p^\mu p^\nu}{2 (p \cdot l)(p \cdot k)
  \left[p \cdot (k + l) \right]} - \frac{{\rm i} \Sigma^{\mu\nu}}{p \cdot (k + l)}
  \right) \right] \, .
\label{Fmomdef} 
\eeqa
Note that we have given the result for a fermion, where $\Sigma^{\mu\nu} = \frac{{\rm i}}{4}
\left[\gamma^\mu,\gamma^\nu\right]$ is the appropriate Lorentz-group spin generator. 
The subscript on the RHS of \eq{Stildedef} indicates than one should truncate the resulting 
expression to include at most one next-to-soft vertex. Correspondingly, one may define 
a next-to-soft jet function
\beq
  \widetilde{\cal J} \Big( p, n \Big) \, = \,  \left.\left\langle 0 \left|
  \Phi_n (\infty, 0) F_{p} (0, -\infty) \right| 0 \right\rangle \, \right|_{\rm NLP} \, .
\label{JdefNE}
\eeq
With these definitions, the non-radiative amplitude reads
\beq
  {\cal A} \left( p_1, p_2 \right) \, = \, \widetilde{\cal H} \left( p_j, n_j \right) \,
  \widetilde{\cal S} \left( p_j \right) \, \prod_{i = 1}^2 \frac{J \left(p_i, n_i \right)}
  {\widetilde{\cal J} \left(p_i,  n_i \right)} \, ,
\label{Adef}
\eeq
as schematically depicted in Fig.~\ref{fig:HJfac2}(a). Just as in \eq{softcolfac}, 
$\widetilde{\cal H}$ is obtained by matching to the full amplitude on the left-hand 
side. It differs from the function ${\cal H}$ appearing in the factorisation formula 
in \eq{softcolfac}, as next-to-soft effects have now been explicitly factored out. 

%%%%%%%%%%%%%%%%%%%%%%%%%

\subsection{Real radiation up to NLP order}
\label{sec:addrad}

\begin{figure}
\begin{center}
  \scalebox{0.95}{\includegraphics{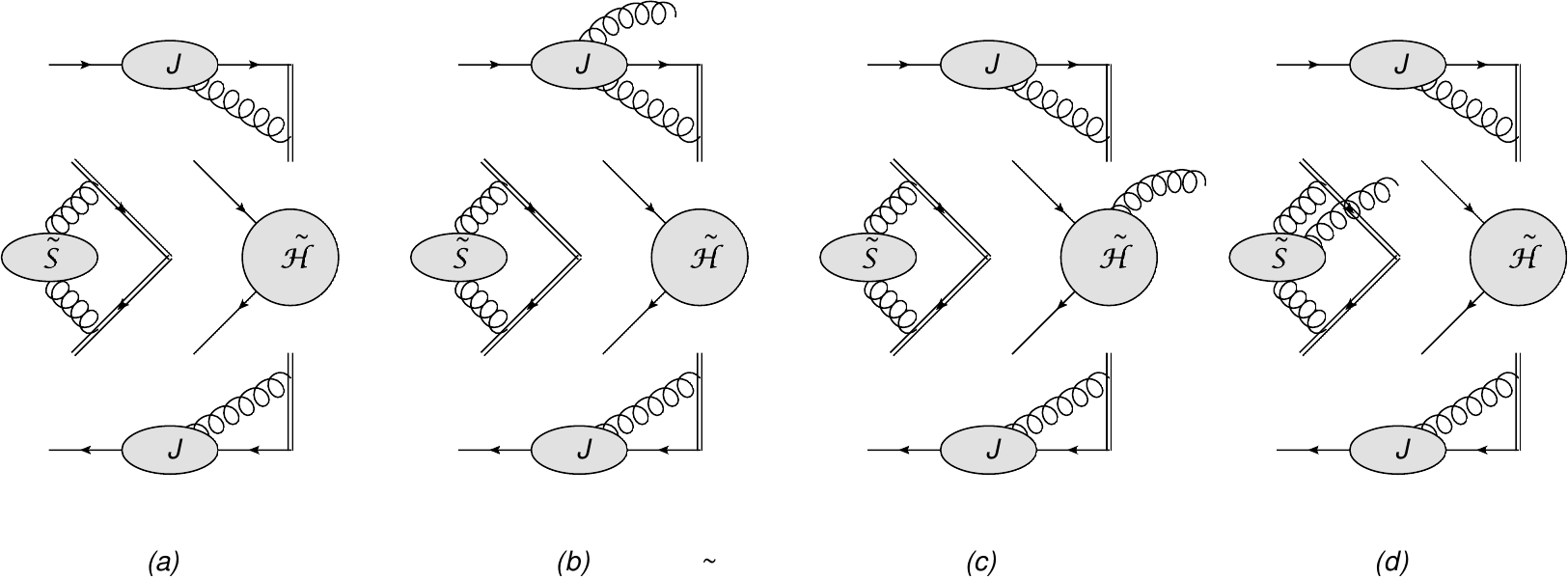}}
  \caption{(a) Schematic factorisation of a two-point amplitude. $\widetilde{\cal H}$ 
  and $\widetilde{\cal S}$ are the hard and next-to-soft functions, and $J$ is a 
  non-radiative jet function; next-to-soft subtractions to $J$ are omitted for simplicity. 
  (b) Emission of a gluon from a jet (to be described by a {\it radiative jet function} 
  $J^\mu$); (c) Emission from the hard function. (d) Emission through a radiative 
  next-to-soft function $\widetilde{S}^\mu$.}
\label{fig:HJfac2}
\end{center}
\end{figure}

Let us now consider adding the radiation of an additional (next-to-)soft gluon to 
the amplitude in \eq{Adef}. The emission of the extra gluon can be assigned to
different factors in the non-radiative amplitude, as shown in Fig.~\ref{fig:HJfac2}(b,c,d).
Proceeding by analogy with the treatment of the abelian case~\cite{DelDuca:1990gz,
Bonocore:2015esa}, we will start by giving a formal definition of the contribution to the
amplitude due to radiation from a jet, say $J(p_1, n_1) \equiv J_1$. We write
\beq
  {\cal A}_{\mu, a}^{J_1} \left(p_1, p_2, k \right) \, = \, \widetilde{\cal H} 
  \left(p_1 - k, p_2, n_j \right) \, \frac{\widetilde{\cal S} \left( p_j \right)}{\prod_{k = 1}^2 
  \widetilde{\cal J} \left(p_k, n_k \right)} \, J_{\mu, a} (p_1, n_1, k) \, J (p_2, n_2) \, ,
\label{AmuJ}
\eeq
where $\mu$ and  $a$ are the Lorentz and the color adjoint indices of the emitted 
gluon, respectively, and we have introduced the {\it radiative jet function}, defined,
as in Refs.\cite{DelDuca:1990gz,Bonocore:2015esa}, by
\beq
  J_{\mu, a} \left( p, n, k \right) u(p) \, = \, \int d^d y \,\, {\rm e}^{ - {\rm i} (p - k) \cdot
  y} \, \left\langle 0 \left| \, \Phi_{n} (\infty, y) \, \psi(y) \, j_{\mu, a} (0) \, 
  \right| p \right\rangle \, .
\label{Jmudef}
\eeq
The crucial issue in generalising the radiative jet function to the non-abelian theory 
is the definition of the non-abelian gauge current $j_{\mu, a}(x)$. First of all, it must 
be a conserved current, $\partial_\mu j^{\mu}_a = 0$, in order for the radiative jet to 
obey the Ward identity
\beq
  k_\mu J^{\mu, a} \left(p, n, k \right) \, = \, g \, {\bf T}^a J \left(p, n \right) \, ,
\label{JmuWard}
\eeq
which is the natural generalisation of the abelian case, and is a necessary ingredient
for the proof of our factorisation formula. Furthermore, we must require that the matrix 
element in \eq{Jmudef} should fully reproduce the relevant terms in the amplitude when 
the (next-to-)soft gluon is radiated from virtual gluons inside the jet. It turns out that 
the standard, textbook definition of the non-abelian Noether current (see, for example,
Ref.~\cite{Weinberg:1996kr}) does not have this property. One must however keep in 
mind that Noether currents are not uniquely defined: in general, it is possible to add
{\it improvement terms} (see, for example, Ref.~\cite{DeWit:1986it}) which do not 
spoil charge conservation but may improve other symmetry properties of the operator.
In our case, we have found that an improvement term indeed exists which reproduces
all relevant terms in diagrams where the (next-to-)soft gluon is emitted though a 
three-gluon vertex. Our choice for the non-abelian current is then
\beq 
  j^\mu_a (x) \, = \, g \, \left\{ - \, \overline{\psi} (x) \, \gamma^\mu \, {\bf T}_a \, \psi (x) 
  \, + \, f_a^{\phantom{a} b c} \Big[ F^{\mu \nu}_c (x) \, A_{\nu \, b} (x)
  + \partial_\nu\left(A^{\mu}_b (x) A^{\nu}_c (x) \right) \Big] \right\} \, ,
\label{nabcurr}
\eeq 
which is indeed conserved, as one can readily verify. Note that the last term in 
\eq{nabcurr} (the `improvement') does not contribute to $\partial_\mu j^\mu_a$.

Our next step is to define an operator matrix element describing (next-to-)soft
real gluon radiation from the soft factor of the non-radiative amplitude, as depicted
in Fig.~\ref{fig:HJfac2}(d). A natural choice is
\beq
  \e_\mu^{* (\lambda)} (k) \, \widetilde{\cal S}^\mu_a \left(p_1, p_2, k \right) \, = \,
  \left. \left\langle k, \lambda, a \left| F_{p_2} \left(\infty, 0 \right) F_{p_1} \left(0, -\infty \right)
  \right| 0 \right\rangle \right|_{\rm NLP} \, .
\label{Smutildedef}
\eeq
This definition is similar to the non-radiative function of \eq{Stildedef}, but one of the 
gluons is now real rather than virtual. A similar quantity is defined at leading power 
in Refs.~\cite{Belitsky:1998tc,Czakon:2009zw,Beneke:2009rj,Idilbi:2009cc,
Czakon:2013hxa}. The radiative next-to-soft function obeys the Ward identity
\beq
  k_\mu \widetilde{S}^\mu \left(p_1, p_2, k \right) \, = \, 0 \, .
\label{SmutildeWard}
\eeq
Following Ref.~\cite{Laenen:2008gt}, the radiative next-to-soft function can be 
shown to exponentiate, and it can be evaluated using {\it next-to-soft webs}, 
generalising the methods used at leading power. These webs can connect all 
hard partons in the process, thus they are not captured by the emission of gluons 
from inside single-parton jets. Nevertheless, there is clearly a double counting 
of contributions between the jets and the next-to-soft function, which is directly 
analogous to the double counting of soft and collinear contributions in \eq{softcolfac}. 
We may correct for this by subtracting from the jet emission contributions defined 
in \eq{AmuJ} their next-to-soft expansion, according to 
\beq
  {\cal A}_{\mu, a}^{J_i} \left(p_i, p_j, k \right) \, \rightarrow \, {\cal A}_{\mu, a}^{J_i} 
  \left(p_i, p_j, k \right) - {\cal A}_{\mu, a}^{\widetilde{\cal J}_i} \left(p_i, p_j, k \right) \, ,
\label{AmucalJ}
\eeq
where the subtraction term on the right-hand side is simply defined as the 
next-to-soft approximation to the full radiative jet function. Given that the 
overlap between the soft and jet functions must be separately gauge-invariant, 
the Ward identity of \eq{SmutildeWard} implies
\beq
  k^\mu {\cal A}^{\widetilde{\cal J}_i}_{\mu, a} \left(p_i, p_j, k \right) \, = \, 0 \, .
\label{JtildemuWard}
\eeq
Having precisely defined the jet and next-to-soft contributions to the radiative 
amplitude in terms of operator matrix elements, and having taken care to subtract 
double counted contributions, the emissions from the hard sub-process, which 
we denote by ${\cal A}_{\mu, a}^{\widetilde{\cal H}}$ and depict in 
Fig.~\ref{fig:HJfac2}(c), are defined by matching to the full radiative amplitude 
${\cal A}_{\mu, a}$. We will discuss their properties in the following subsection.

%%%%%%%%%%%%%%%%%%%%%%%%%

\subsection{Derivation of the non-abelian factorisation formula}
\label{sec:nonabelfac}

Combining the above ingredients gives a total radiative amplitude 
\beqa
  {\cal A}_{\mu, a} \left(p_i, k \right) & = & {\cal A}^J_{\mu, a} \left(p_i, n_i, k \right) - 
  {\cal A}^{\widetilde{\cal J}}_{\mu, a} \left(p_i, n_i, k\right) + 
  {\cal A}^{\widetilde{\cal H}}_{\mu, a} \left(p_i, n_i, k \right) \nonumber \\
  && + \, \widetilde{\cal S}_{\mu, a} \left( p_i, k \right) \, 
  \left( \widetilde{{\cal H}} \left(p_i, n_i \right) \prod_{i=1}^2 
  \frac{J \left(p_i, n_i \right)}{\tilde{\cal J} \left(p_i, n_i\right)} \right) \, ,
\label{Amudef}
\eeqa
where we defined
\beq 
  {\cal A}_{\mu, a}^J \left(p_1, n_1, p_2, n_2, k \right) \, \equiv \, 
  \sum_{i = 1}^2 {\cal A}_{\mu, a}^{J_i} \left(p_i, n_i, k \right) \, ,
\label{sumJi}
\eeq
and similarly for ${\cal A}_{\mu, a}^{\widetilde{\cal J}}$. The complete radiative 
amplitude in \eq{Amudef} satisfies the Ward identity
\beq
  k^\mu {\cal A}_\mu^a \left(p_1, p_2, k \right) \, = \, 0 \, ,
\label{Wardamp}
\eeq
which, together with Eqs.~(\ref{JmuWard}, \ref{SmutildeWard}, \ref{JtildemuWard}) 
implies the relation
\beq
  k^\mu \, {\cal A}^{\widetilde{\cal H}}_{\mu, a} \, = \, - \, k^\mu \, 
  {\cal A}^J_{\mu, a} \, .
\label{Ward}
\eeq
Taylor expanding \eq{AmuJ} and using \eq{Ward} and colour conservation, in the
form $\sum_i {\bf T}_i = 0$, one finds
\beq
  {\cal A}^{\widetilde{\cal H}}_{\mu, a} \left(p_j, n_j, k \right) \, = \, \sum_{i = 1}^2 g \, 
  {\bf T}^a_i \left[ \frac{\partial} {\partial p_i^\mu} \widetilde{\cal H} \left( p_j, n_j 
  \right) \right] \widetilde{\cal S} \left(p_j \right) \prod_{k = 1}^2 
  \frac{J (p_k,n_k)}{\widetilde{\cal J}(p_k,n_k)} \, .
\label{AHexpand2}
\eeq
In order to characterise the radiative jet functions it is convenient, as in
Refs.~\cite{DelDuca:1990gz,Bonocore:2015esa} to introduce polarisation
tensors~\cite{Grammer:1973db}
\beq
  \eta^{\mu \nu} \, = \, G^{\mu \nu}_i + K^{\mu \nu}_i \, , \qquad
  K^{\mu \nu}_i \, = \, \frac{(2 p_i - k)^\nu}{2 p_i \cdot k - k^2} \, k^\mu \, ,
\label{KGdef}
\eeq
so that the total radiative amplitude is given by the sum of ``$K$-polarised'' and 
``$G$-polarised'' gluons. Considering first the K-projection of the jet contribution to 
the  amplitude, and using Eqs.~(\ref{AmuJ}, \ref{JmuWard}, \ref{JtildemuWard}),
we find
\beqa
  \sum_{i = 1}^2 \left( {\cal A}_{\nu, a}^{J_i} - {\cal A}_{\nu, a}^{\widetilde{\cal J}_i} 
  \right) \, K_i^{\nu \mu} & = & \sum_{i = 1}^2 g \, {\bf T}_{a, i} \left[ 
  \frac{(2 p_i - k)^\mu}{2 p_i \cdot k - k^2} \, \widetilde{\cal H} \left( p_j, n_j \right)
  \widetilde{\cal S} \left( p_j \right) \prod_k \frac{J(p_k,n_k)}
  {\widetilde{\cal J}(p_k,n_k)} \right. \nonumber \\
  & & \left. - \left( K_i^{\nu \mu} \frac{\partial}{\partial p_i^\nu} \, 
  \widetilde{\cal H} \left( p_j, n_j \right) \right) \widetilde{\cal S} (p_j) 
  \prod_k \frac{J(p_k,n_k)}{\widetilde{\cal J}(p_k,n_k)} \right] \, ,
\label{KAJ}
\eeqa
where we have again Taylor expanded in $k$. The emission of a $G$-gluon from a 
jet, on the other hand, is given by
\beq
  \sum_{i = 1}^2 {\cal A}^{J_i}_{\nu, a} \, G^{\nu \mu}_i \, = \, \sum_{i = 1}^2 \, 
  G^{\nu\mu}_i \, \widetilde{\cal H} \left( p_j, n_j \right) \,
  J_{\nu, a} \left( p_i, n_i, k \right) \frac{\widetilde{\cal S} (p_j)}{\prod_{k = 1}^2 
  \tilde{\cal J}(p_k,n_k)} \, \prod_{j \neq i} J(p_j, n_j) \, ,
\label{Gjet}
\eeq
where we set $k \rightarrow 0$ in the hard function, retaining only the leading 
term in its Taylor expansion, owing to the fact that the $G$ tensor acts on terms 
proportional to $p_i^\mu$ to make them ${\cal O}(k)$~\cite{DelDuca:1990gz,
Bonocore:2015esa}. Combining this with the $K$-gluon emissions, with the 
$G$-gluon contribution from the subtraction term, and with emissions from 
the hard function, as given by \eq{AHexpand2}, we find that the total amplitude, 
to the required accuracy, becomes\footnote{For convenience, in what follows, 
we have chosen to keep terms that vanish due to the on-shell condition for 
the emitted gluon, $k^2=0$, and due to the physical polarisation condition, 
$k_\mu\e^\mu(k)=0$.}
\beqa
  {\cal A}_{\mu, a} \left(p_j, k\right) & = & \sum_{i = 1}^2 \Bigg\{ \left[ 
  \frac12 \, \widetilde{\cal S}_{\mu, a} (p_j, k) \, \widetilde{\cal H} \left( p_j, n_j \right)
  + g \, {\bf T}^a_i G^{\nu \mu}_i \, \left( \frac{\partial}{\partial p_i^\nu}
  \widetilde{\cal H} \left( p_j, n_j \right) \right) \widetilde{\cal S} (p_j) \right] 
  \nonumber \\
  & & \hspace{-12mm} \times \, \prod_{j} \frac{J(p_j,n_j)}{\widetilde{\cal J}(p_j,n_j)}
  + \widetilde{\cal H} \left( p_j, n_j \right) \widetilde{\cal S} (p_j) \,
  \frac{J_{\mu, a} \left(p_i, n_i, k\right)}{\widetilde{\cal J}(p_i,n_i)}
  \prod_{j \neq i} \frac{J(p_j,n_j)}{\widetilde{\cal J}(p_j,n_j)}
  - {\cal A}^{\widetilde{\cal J}_i}_{\mu, a} \Bigg\} \, ,
\label{factorisation2}
\eeqa
where the factor of $1/2$ in the first term is due to the fact that we have placed 
this inside the sum over hard particles, for brevity. 

Reconstructing the expression for the non-radiative amplitude given in \eq{Adef}, 
we may express \eq{factorisation2} as
\beqa
  {\cal A}_{\mu, a} \left(p_j, k\right) & = & \sum_{i = 1}^2 \Bigg\{ 
  \Bigg[ \frac12 \, \frac{\widetilde{\cal S}_{\mu, a} (p_j, k)}{\widetilde{\cal S} (p_j)}
  + g \, {\bf T}_{i, a} G^\nu_{i, \mu} \, \frac{\partial}{\partial p_i^\nu} 
  + \frac{J_{\mu, a} \left(p_i, n_i, k\right)}{J(p_i,n_i)} 
  \label{NEfactor_nonabel} \\
  & & \qquad \quad - \, g \, {\bf T}_{i, a} \, G_{i, \mu}^\nu \, \frac{\partial}{\partial p_i^\nu}
  \log \left( \frac{J(p_i,n_i)}{\widetilde{\cal J}(p_i,n_i)} \right) \Bigg]
  {\cal A} (p_j) - {\cal A}^{\widetilde{\cal J}_i}_{\mu, a} (p_j, k) 
  \Bigg\} \, , \nonumber 
\eeqa
where we have used \eq{Adef} to replace derivatives of the hard interaction 
with those acting on the full non-radiative amplitude and jet functions. 
\eq{NEfactor_nonabel} is our final non-abelian factorisation formula, 
capturing all NLP contributions near threshold. A few comments are in order.
\begin{itemize}
\item The first two terms in \eq{NEfactor_nonabel} contain next-to-eikonal 
webs, composed of generalised Wilson lines~\cite{Laenen:2008gt}, dressing 
the non-radiative amplitude, together with a derivative operator. These terms 
provide a non-abelian version of the original Low's theorem (in the absence 
of collinear enhancements).   
\item The remaining terms in square brackets organise emissions from jet 
functions, generalising to the non-abelian theory the results of~\cite{DelDuca:1990gz,
Bonocore:2015esa}.
\item
The last term in \eq{NEfactor_nonabel} corrects the radiative jet factors for
the double counting of contributions between the radiative jets and next-to-soft 
functions.
\item \eq{NEfactor_nonabel} describes the amplitude stripped of external 
wave functions for the hard partons. To build a cross section, these must be 
reinstated, as in \eq{Mdef}, noting that the derivative in \eq{NEfactor_nonabel} 
does not act on the wave functions. 
\end{itemize}
As discussed in Refs.~\cite{Bonocore:2015esa,Bonocore:2014wua},
considerable simplifications occur in \eq{NEfactor_nonabel} upon choosing 
the auxiliary vectors $n_i$ to be null, $n_i^2 = 0$. In this case, one may work 
in a renormalisation scheme such that the non-radiative soft and jet functions 
are unity, to all orders in perturbation theory. \eq{NEfactor_nonabel} then 
becomes
\beq
  {\cal A}_{\mu, a} (p_j, k) \, = \, \sum_{i = 1}^2 \left( \frac12 \, 
  \widetilde{\cal S}_{\mu, a} (p_j, k) + g \, {\bf T}_{i, a} \, G^\nu_{i, \mu} 
  \, \frac{\partial}{\partial p_i^\nu} + J_{\mu, a} \left(p_i, n_i, k\right)
  \right) {\cal A} (p_j) - {\cal A}^{\widetilde{\cal J}}_{\mu, a} (p_j, k) \, .
\label{NEfactor_nonabel2}
\eeq
As in Refs.~\cite{Bonocore:2015esa,Bonocore:2014wua}, in the detailed 
calculations below we will further make the specific choice of reference vectors
\beq
  n_1 \, = \, p_2 \, , \quad n_2 \, = \, p_1 \, ,
\label{nchoices}
\eeq
whose interpretation is that $n_i$ is the anti-collinear direction associated with 
$p_i$. This is physically motivated by the fact that $p_1$ and $p_2$ are the only
momenta in the problem at hand, and it allows to make direct contact with the
method-of-regions calculation of Ref.~\cite{Bonocore:2014wua}.

%%%%%%%%%%%%%%%%%%%%%%%%%%%%%%%%%%%%%%%%%%%%%

\section{The non-abelian radiative jet function}
\label{sec:jetemit}

\begin{figure}
\begin{center}
  \scalebox{0.8}{\includegraphics{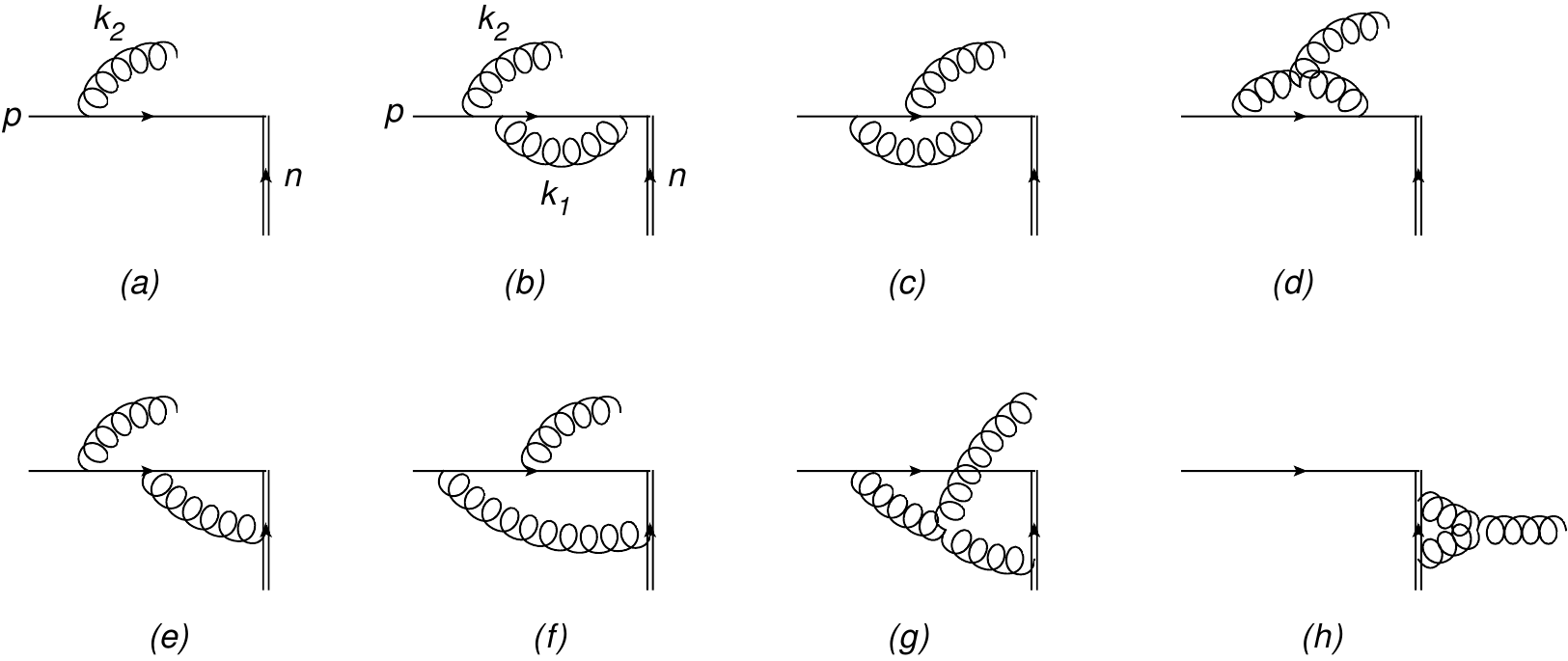}}
  \caption{Diagrams contributing to radiative quark jet function: (a) tree level; 
  (b)--(h) one-loop.} 
\label{fig:Jmudiags}
\end{center}
\end{figure}

\begin{figure}
\begin{center}
  \scalebox{0.7}{\includegraphics{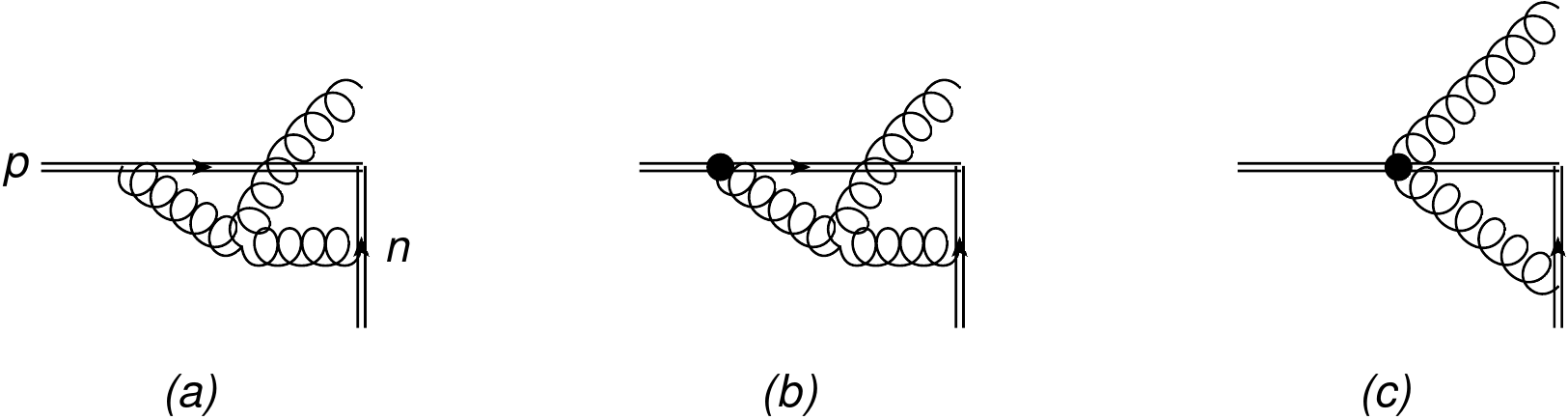}}
  \caption{Example diagrams for the next-to-soft radiaive jet function, where the 
  $p$ leg has been replaced by a generalised Wilson line, and $\bullet$ denotes 
  a next-to-soft emission vertex, arising from \eq{Fmomdef}.}
\label{fig:calJdiags}
\end{center}
\end{figure}

Before testing \eq{NEfactor_nonabel2} in Drell-Yan production, we must first calculate 
the non-abelian radiative quark jet function, defined in \eq{Jmudef}. To perform a test
at NNLO, we need to compute $J_{\mu, a}$ at one loop, which we do for null $n$, in 
order to use the result in \eq{NEfactor_nonabel2}. Relevant Feynman diagrams are shown in Fig.~\ref{fig:Jmudiags}. Defining the perturbative coefficients of 
$J_{\mu, a}$ via
\beq
  J^a_\mu \left( p, n, k \right)  \, = \, g \, {\bf T}^a \, \sum_{n = 0}^{\infty}
  \left(\frac{\alpha_s}{4 \pi}\right)^n J^{(n)}_\nu \left(p, n, k \right) \, ,
\label{pertjmu}
\eeq
the diagram of Fig.~\ref{fig:Jmudiags}(a) gives
\beq
  J^{(0)}_\mu \left(p, n, k \right) \, = \, - \, \frac{p_\mu}{p \cdot k} + \frac{k_\mu}{2 p \cdot k}
  - \frac{{\rm i} \, k_{\alpha} \Sigma^\alpha_\mu}{p \cdot k} \, .
\label{Jnu0}
\eeq
One-loop diagrams are shown in Fig.~\ref{fig:Jmudiags}(b)--(h). Notice that we are
computing the bare $J_{\mu, a}$, as required in \eq{NEfactor_nonabel2}, following
the discussion in Ref.~\cite{Bonocore:2015esa}. We can then use the fact that the 
integral for (h) is scaleless for null $n$ and thus vanishing. Similarly, we have omitted
external-leg vacuum polarisations dressing the tree-level diagram. The result can be
cast in the form
\beq
  J^{(1)}_\mu \, = \, (- 2 p \cdot k)^{-\e} \left[ C_F J^{(1)}_{\mu, F} + C_A
  J^{(1)}_{\mu, A, {\rm coll.}} \right] + \left( \frac{2 p \cdot n}{(- 2 p \cdot k)(- 2 n \cdot k)}
  \right)^\e C_A \, J^{(1)}_{\mu, A, {\rm soft}} \, .
\label{Jmutot}
\eeq
Defining the kinematic variables
\beq
  t \, = \, - 2 p \cdot k \, , \quad  n_p^\mu \, = \, \frac{n^\mu}{2 n \cdot p} \, ,
  \quad  n_k^\mu \, = \, \frac{n^\mu}{2 n \cdot k} \, , \quad r \, = \,  
  \frac{n \cdot k}{n \cdot p} \, ,
\label{mandieslor}
\eeq
we find that the coefficients in \eq{Jmutot} can be written as
\beqa
  J^{(1)}_{\mu, F} & = & \left(1 + 2 \e \right) \frac{\slash{k}}{t} \, \gamma^\mu + 
  \left(2 + 6 \e \right) \slash{n}_p \gamma^\mu + \left[ \frac{2}{\e} - 2  - \e \left( 8
  + \zeta_2 \right) \right] \frac{k^\mu}{t} 
  \nonumber \\
  & & + \, \left[ \frac{4}{\e} + 4 + 2 \e \left( 2 - \zeta_2 \right) \right] n_p^\mu + 
  \left[ \frac{4}{\e} + 8 - 2 \e \left(- 8 + \zeta_2 \right) \right] 
  \frac{r}{t} \, p^\mu - 4 \left(1 + 3 \e \right) \frac{\slash{k} \slash{n}_p}{t} \, p^\mu \, ; 
  \nonumber \\
  & & \phantom{abcd} 
  \nonumber \\
  J^{(1)}_{\mu, A, {\rm coll.}} & = & - \left( 1 + 2 \e \right) \frac{\slash{k}}{t} \, 
  \gamma^\mu +  \left[\frac{1}{\e} + 1 + \e \left(1 - \frac{\zeta_2}{2} \right) \right]
  \slash{n}_p \, \gamma^\mu - 4 n_p^\mu 
  \nonumber \\ 
  & & - \left[- \frac{2}{\e} + 2 + \e \left(- 2 + \zeta_2 \right) \right] \frac{r}{t} \, p^\mu
  + \left[ - \frac{2}{\e} - 2 + \e \left(- 2 + \zeta_2 \right) \right] \frac{\slash{k} 
  \slash{n}_p}{t} \, p^\mu
  \nonumber \\ 
  & & + \left[- \frac{1}{\e^2} - \frac{3}{\e} - 3 + \frac{\zeta_2}{2} + 
  \e \left(- 4 + \frac{3}{2} \zeta_2 + \frac{7}{3} \zeta_3 \right) \right]
  \frac{k^\mu}{t} \, ; 
  \nonumber \\
  & & \phantom{abcd} 
  \nonumber \\
  J^{(1)}_{\mu, A, {\rm soft}} & = & \left[ - \frac{2}{\e} - 4 + \e \left( - 8
  - \zeta_2 \right) \right] \slash{n}_p \, \gamma^\mu
  + \Bigg\{ - \left[\frac{2}{\e} +  4 + \e \left( 8 + \zeta_2 \right) \right] n_k^\mu
  \nonumber \\ 
  & & + \left[ \frac{2}{\e} + 4 + \e \left( 8 + \zeta_2 \right) \right] \frac{p^\mu}{t} 
  - \left[ \frac{1}{\e^2} + \frac{2}{\e} + 4 + \frac{\zeta_2}{2} + 
  \e \left( 8 + \zeta_2 - \frac{7}{3} \zeta_3 \right) \right] \frac{k^\mu}{r t}
  \Bigg\} \, \slash{k} \slash{n}_p 
  \nonumber \\
  & & + \left[ \frac{1}{\e^2} + \frac{2}{\e} + 4 + \frac{\zeta_2}{2} + 
  \e \left( 8 + \zeta_2 - \frac{7}{3} \zeta_3 \right) \right] \, \frac{k^\mu}{t} 
  + \left(\frac{1}{\e^2} + \frac{\zeta_2}{2} - \frac{7}{3} \e \, \zeta_3 \right)
  \frac{\slash{k}}{t} \, \gamma^\mu
  \nonumber \\
  & & + \left\{- \frac{2}{\e^2} - \zeta_2 + \frac{14}{3} \e \, \zeta_3 -
  \left[ \frac{2}{\e} + 4 + \e \left( 8 + \zeta_2 \right) \right] r \right\}\frac{p^\mu}{t} 
  \nonumber \\
  & & + \left[ \frac{2}{\e} + 4 + \e \left( 8 + \zeta_2 \right) - \left(\frac{2}{\e^2} 
  + \zeta_2 - \frac{14}{3} \e \, \zeta_3 \right) \frac{1}{r} \right] \, n_p^\mu \, .
\label{Jmutot2}
\eeqa
The first two terms in \eq{Jmutot} are accompanied by a factor $(2 p \cdot k)^{-\e}$, 
corresponding to the collinear scale associated with radiation from a 
jet~\cite{Bonocore:2014wua,Bonocore:2015esa}. The third term in \eq{Jmutot}, 
on the other hand, contains a different ratio of scales involving the auxiliary vector 
$n$. Note that for the choices in \eq{nchoices} the ratio for both jets becomes
\beq
  \left( \frac{2 p \cdot n}{( - 2 p \cdot k)( - 2 n \cdot k)} \right)^\e \, \rightarrow \, 
  \left(\frac{2 p_1 \cdot p_2}{( - 2 p_1 \cdot k)( - 2 p_2 \cdot k)} \right)^\e \, .
\label{softratio}
\eeq
This is the same dependence arising in (next-to-)soft webs connecting both 
external partons (shown for example in Fig.~\ref{fig:HJfac2}(d)). Terms with 
this scale dependence thus constitute the double counting of overlapping 
(next-to-)soft and collinear regions for the virtual gluon momentum, to be 
removed by the subtraction term ${\cal A}^{\widetilde{\cal J}}_{\mu, a}$. In 
our present calculation, one may interpret this overlap diagrammatically
by defining a {\it next-to-soft radiative jet function} $\widetilde{\cal J}_{\mu, a}
(p, k, n)$. This function appears in the subtraction term ${\cal A}^{\widetilde{\cal 
J}}_{\mu, a}$ instead of the full radiative jet function used in the definition 
of ${\cal A}_{\mu, a}^{J_i}$, \eq{AmuJ}. By analogy with \eq{AmuJ}, we then write
\beq
  {\cal A}_{\mu, a}^{\widetilde{\cal J}_1} \left(p_1, p_2, k \right) \, = \, \widetilde{\cal H} 
  \left(p_1, p_2, n_j \right) \, \frac{\widetilde{\cal S} \left( p_j \right)}{\prod_{k = 1}^2 
  \widetilde{\cal J} \left(p_k, n_k \right)} \, \widetilde{\cal J}_{\mu, a} (p_1, n_1, k) \, 
  J (p_2, n_2) \, .
\label{AmuJns}
\eeq
The function $\widetilde{\cal J}_{\mu, a}$ can be obtained from the diagrams for 
the full radiative jet, by replacing the emission vertices on the $p$ leg with the soft 
or next-to-soft Feynman rules arising from \eq{Fmomdef}, and including at most 
one next-to-soft vertex. At tree-level (using the normalisation of \eq{pertjmu}) one 
simply finds $\widetilde{\cal J}^{(0)}_\mu (p, n, k) =  J^{(0)}_\mu (p, n, k)$.  At the
one-loop level, one encounters diagrams such as those in Fig.~\ref{fig:calJdiags}: 
in fact, only the diagrams in Fig.~\ref{fig:calJdiags}(a) and (b) are non-vanishing,
By analogy with \eq{Jmutot}, one can write the result in the form
\beq
  \widetilde{\cal J}^{(1)}_\mu \, = \, \left( - 2 p \cdot k \right)^{- \e} \left[ C_F 
  \widetilde{\cal J}^{(1)}_{\mu, F} + C_A \widetilde{\cal J}^{(1)}_{\mu, A, {\rm coll.}} 
  \right] + \left( \frac{2 p \cdot n}{( - 2 p \cdot k)(- 2 n \cdot k)} \right)^\e C_A \, 
  \widetilde{\cal J}^{(1)}_{\mu, A, {\rm soft}} \, ,
\label{Jmutildetot}
\eeq
and one finds that
\beq
  \widetilde{\cal J}^{(1)}_{\mu, F} \, = \, \widetilde{\cal J}^{(1)}_{\mu, A, {\rm coll.}} 
  \, = \, 0 \, , \quad
  \widetilde{\cal J}^{(1)}_{\mu, A, {\rm soft}} \, = \, J^{(1)}_{\mu, A, {\rm soft}} \, ,
\label{Jmutildetot2}
\eeq
so that the next-to-soft radiative jet function reproduces precisely the third 
term in \eq{Jmutot}: subtracting it from the full jet leaves only collinear 
contributions, as required.

According to \eq{NEfactor_nonabel2}, for the complete result one also needs the
radiative next-to-soft function $\widetilde{\cal S}_\mu$ at one-loop. The relevant 
diagrams are similar to those entering the next-to-soft radiative jet function. The 
leading power soft diagrams can be obtained simply by relabelling $p \rightarrow p_1$, 
$n \rightarrow p_2$. For the next-to-soft contribution, there are two sets of diagrams: 
those where the next-to-soft emission vertex is on leg $p_1$, and those where it is 
on leg $p_2$. One may then write $\widetilde{\cal S}^\mu = \widetilde{\cal S}^\mu_{\rm E} 
+ \widetilde{\cal S}^\mu_{\rm NE}$, with
\beq
  \widetilde{S}_{\rm E}^\mu \, = \, \widetilde{\cal J}^\mu_{\rm E} \big|_{p \rightarrow p_1,
  n \rightarrow p_2} \, , \quad
  \widetilde{S}_{\rm NE}^\mu \, = \, \widetilde{\cal J}^\mu_{\rm NE} \big|_{p \rightarrow p_1,
  n \rightarrow p_2} + \widetilde{\cal J}^\mu_{\rm NE} \big|_{p \rightarrow p_2,
  n \rightarrow p_1} \, ,
\label{SEdef}
\eeq
where the subscripts E and NE refer to eikonal and next-to-eikonal contributions respectively. 

%%%%%%%%%%%%%%%%%%%%%%%%%%%%%%%%%%%%%%%%%%%%%

\section{Application to Drell-Yan production}
\label{sec:DY}

We now have all ingredients to verify \eq{NEfactor_nonabel2} in the Drell-Yan process
\beq
  q(p_1) + \bar{q} (p_2) \rightarrow V^*(Q) \, ,
\label{DYLO}
\eeq
where $q$ and $\bar{q}$ denote a quark and antiquark respectively, $V^*$ an off-shell 
vector boson, and arguments label 4-momenta. At cross-section level, all LP and NLP
threshold logarithms are associated with real emission of soft or next-to-soft gluons; 
virtual gluons, however, can be hard and collinear, thus loop corrections test all 
ingredients in \eq{NEfactor_nonabel}. As usual, one defines the threshold variable 
$z \, = \, Q^2/s$, representing the fraction of available energy carried by the 
final state vector boson; the threshold limit then corresponds to $z \rightarrow 1$. 
The K-factor at fixed order in perturbation theory is defined by
\beq
  K^{(n)}(z) \, = \, \frac{1}{\sigma^{(0)}} \, \frac{d \sigma^{(n)}(z)}{dz} \, ,
\label{Kndef}
\eeq
with $\sigma^{(n)}$ the $n$-loop cross section. As was the case in~\cite{Bonocore:2015esa,
Bonocore:2014wua}, the first non-trivial test of \eq{NEfactor_nonabel} is to reproduce
the real-virtual contribution to the NNLO K-factor. To do so, we need the tree-level and 
one-loop amplitudes with one real emission. Using \eq{NEfactor_nonabel}, we find
\beq
  {\cal A}_{\mu, a}^{(0)} \, = \, g \left[- {\bf T}_{1, a} \, {\cal A}^{(0)} 
  \left( \frac{p_{1, \mu}} {p_1 \cdot k} + \frac{{\rm i } k_\alpha 
  \Sigma^\alpha_{\phantom{\alpha} \mu}}{p_1\cdot k}\right)
  + {\bf T}_{2, a} \left( \frac{p_{2, \mu}}{p_2 \cdot k} + 
  \frac{{\rm i} k_\alpha \Sigma^\alpha_{\phantom{\alpha} \mu}}{p_2
  \cdot k}\right) {\cal A}^{(0)} \right] \, ,
\label{Ardef}
\eeq
where ${\cal A}^{(0)}$ is the leading order non-radiative amplitude, stripped of external 
spinors, and we have used the tree-level radiative jet function in \eq{Jnu0}, as well as 
the physical polarisation condition $k_\mu \e^\mu(k) = 0$. We have also defined colour 
generators ${\bf T}^a_{1,2}$ acting on the $p_1$, $p_2$ legs respectively.
For the one-loop amplitude, we use \eq{NEfactor_nonabel2}, which gives
\beq
  {\cal A}_{\mu, a}^{(1)} \, = \, \sum_{i = 1}^2 \bigg\{ \left[ \frac12 \, 
  \widetilde{\cal S}^{(0)}_{\mu, a} + g {\bf T}_{i, a} G^\nu_{i, \mu} \, 
  \frac{\partial}{\partial p_i^\nu} + J^{(0)}_{\mu, a} - \widetilde{\cal J}^{(0)}_{\mu, a}
  \right] {\cal A}^{(1)} + \left[ \frac12 \, \widetilde{\cal S}^{(1)}_{\mu, a} + 
  J^{(1)}_{\mu, a} - \widetilde{\cal J}^{(1)}_{\mu, a} \right] {\cal A}^{(0)} \bigg\} \, ,
\label{Arvdef}
\eeq
where in the second term we have used the fact that the derivative of the 
non-radiative tree-level amplitude vanishes~\cite{Bonocore:2015esa}. \eq{Arvdef} 
can be further simplified by noting that, at tree level, the next-to-soft function 
contribution precisely cancels the next-to-soft radiative jet contribution.The only 
missing ingredient at this point is the derivative of the one-loop non-radiative 
amplitude, which was already derived in Ref.~\cite{Bonocore:2015esa}. For
example, the contribution from the $p_1$ leg is given by
\beq
  G^{\nu \mu}_1 \, \frac{\partial{\cal A}^{(1)}}{\partial p_1^\nu} \, = \,
  - \, \frac{\e}{p_1 \cdot p_2} \left( - \, p_1^\mu
  + \frac{p_2 \cdot k}{p_1 \cdot k} \, p_2^\mu \right) {\cal A}^{(1)} \, .
\label{Gderivres}
\eeq
It is straightforward to assemble all the ingredients~\footnote{Results for the 
non-radiative amplitude up to one-loop, as well as parametrisations of phase 
space integrals in the present notation, may be found in Ref.~\cite{Bonocore:2015esa}.
In the result we present, as was done in Ref.~\cite{Bonocore:2015esa}, we 
neglect terms involving transcendental constants for brevity, and we do not 
include $\delta$-function terms, which mix with the fully virtual two-loop 
contribution.}, to compute the full real-virtual contribution NNLO K-factor.
We find
\beqa
  K_{\rm rv}^{(2)}(z) & = & \left( \frac{\as}{4 \pi} \right)^2 
  \Bigg\{ \, C_F^2 \, \bigg[ \frac{32{\cal D}_0(z) - 32}{\eps^3}
  + \frac{- 64{\cal D}_1(z) + 48 {\cal D}_0(z) + 64 L(z) - 96}{\eps^2}
  \nonumber \\ 
  & & \hspace{-10mm} + \, \frac{64{\cal D}_2 (z) - 96 {\cal D}_1 (z) + 
  128 {\cal D}_0 (z) - 64 L^2 (z) + 208 L(z) - 196}{\eps} - 
  \frac{128}{3} {\cal D}_3 (z)
  \nonumber \\
  & & \hspace{-10mm} + \, 96 {\cal D}_2 (z) - 256 {\cal D}_1(z) + 
  256 {\cal D}_0(z) + \frac{128}{3} L^3(z) - 232 L^2(z) + 412 L(z) 
  - 408 \bigg]
  \nonumber \\
  & & \hspace{-10mm} + \, C_A C_F \, \bigg[ \frac{8 {\cal D}_0(z) - 8}{\eps^3}
  + \frac{- 32{\cal D}_1(z) + 32 L(z) - 16}{\eps^2}  
  + \frac{64 {\cal D}_2(z) - 64 L^2(z) + 64 L(z) + 20}{\eps}
  \nonumber \\
  & & \hspace{-10mm} - \, \frac{256}{3}{\cal D}_3 (z) + \frac{256}{3} L^3(z) - 
  128 L^2(z) - 60 L(z) + 8 \bigg] \bigg\} \, ,
\label{NNLOTot}
\eeqa
where
\beq
  {\cal D}_i (z) \, = \, \left( \frac{\log^i (1 - z)}{(1 - z)} \right)_+ \, ,
  \quad L(z) \, = \, \log(1 - z) \, .
\label{Ddef}
\eeq
For comparison with the exact two-loop calculation, we note that the real-virtual 
contribution to the NNLO K-factor is not separately available in the
literature~\cite{Hamberg:1991np,Matsuura:1991pc}. We have performed an 
independent calculation of this result, similar to the one carried out for the 
abelian-like contributions in Ref.~\cite{Bonocore:2015esa}. We find that
\eq{NNLOTot} reproduces exactly the full NNLO result, when the latter is
truncated to NLP in $(1 - z)$, including non-logarithmic contributions.

%%%%%%%%%%%%%%%%%%%%%%%%%%%%%%%%%%%%%%%%%%%%%

\section{Double real emission contributions}
\label{sec:tworeal}

\begin{figure}
\begin{center}
  \scalebox{0.8}{\includegraphics{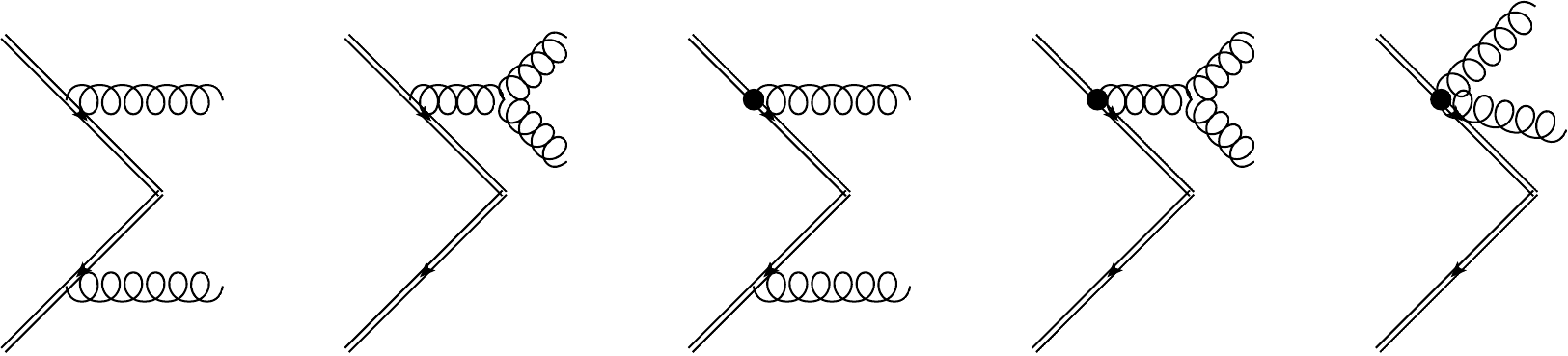}}
  \caption{Sample diagrams contributing to the double radiative next-to-soft function, 
  where $\bullet$ denotes a next-to-eikonal Feynman rule, and all other couplings to
  the external lines are eikonal.}
\label{fig:SmunudiagsNEb}
\end{center}
\end{figure}

In \secn{sec:DY}, we have focused on a single additional gluon emission dressing the 
non-radiative amplitude. Although a full factorisation formula for multiple emissions is 
beyond the scope of this paper, we can nevertheless obtain the double-real emission 
contributions to Drell-Yan production at NNLO by noting that all purely real-emission
near-threshold contributions are (next-to-)soft in nature, with no hard collinear terms. 
This fact was already exploited in Ref.~\cite{Laenen:2010uz}, where next-to-soft 
Feynman rules were employed to compute the abelian part of the NNLO K-factor for 
double real emission. That calculation can easily be reproduced and generalised 
to the full non-abelian theory in the present framework. In essence, all relevant terms 
can be obtained by dressing the Born amplitude with (next-to-)soft webs. Formally, by
analogy with \eq{Smutildedef}, we may define a double radiative next-to-soft function 
according to
\beq
  \e_{\mu, \lambda_1}^{*} (k_1) \, \e_{\nu, \lambda_2}^{*}(k_2) \,
  \widetilde{\cal S}^{\mu \nu} (p_1, p_2, k_1, k_2) \, = \, 
  \left\langle k_1, \lambda_1; k_2, \lambda_2 \left| F_{p_2} (\infty, 0) F_{p_1} 
  (0, - \infty) \right| 0 \right\rangle \big|_{\rm NLP} \, .
\label{Smunutildedef}
\end{equation}
A sampling of soft and next-to-soft diagrams resulting from this definition are
shown in Fig.~\ref{fig:SmunudiagsNEb}.
We have evaluated all diagrams using the next-to-soft Feynman rules arising from 
\eq{Fmomdef}, and we have integrated over the three-body phase space as in 
Refs.~\cite{Hamberg:1991np,Laenen:2010uz}. The result for the double real emission 
contribution to the NNLO K factor is
\beqa
  K^{(2)}_{\rm rr} (z) & = &  \left( \frac{\as}{4 \pi} \right)^2 
  \Bigg\{ \, C_F^2 \, \Bigg[ - \frac{32 {\cal D}_0 (z) - 32}{\e^3} + 
  \frac{128 {\cal D}_1(z) - 128 L(z) + 80}{\e^2}
  \nonumber \\
  & &  - \, \frac{256 {\cal D}_2 (z) - 256 L^2 (z) + 320 L(z)}{\e}
  + \frac{1024}{3}{\cal D}_3 (z) - \frac{1024}{3} L^3 (z) + 640 L^2 (z) \Bigg]
  \nonumber \\ 
  & & + \, C_A C_F \, \Bigg[ - \frac{8 {\cal D}_0 (z) - 8}{\e^3}
  + \frac{1}{\e^2} \left( 32{\cal D}_1 (z) - \frac{44}{3}{\cal D}_0 (z) - 32 L (z) + 
  \frac{92}{3} \right)
  \nonumber \\ 
  & & + \, \frac{1}{\e} \left( - 64 {\cal D}_2 (z) + \frac{176}{3} {\cal D}_1 (z) - 
  \frac{268}{9} {\cal D}_0 (z) + 64 L^2 (z) - \frac{368}{3} L (z) + \frac{520}{9} \right)
  \nonumber \\ 
  & & + \, \frac{256}{3} {\cal D}_3 (z) - \frac{352}{3} {\cal D}_2 (z) + \frac{1072}{9} {\cal D}_1 (z) 
  - \frac{1616}{27} {\cal D}_0 (z) 
  \nonumber \\
  & & - \, \frac{256}{3} L^3 (z) + 
  \frac{736}{3} L^2 (z) - \frac{2080}{9} L (z) + \frac{2912}{27}  \Bigg]
  \nonumber \\ 
  & &  + \, n_f C_F \, \Bigg[ \frac{8 {\cal D}_0 (z) - 8}{3 \e^2}
  + \frac{1}{\e} \left( -\frac{32}{3} {\cal D}_1 (z) + \frac{40}{9} {\cal D}_0 (z) 
  + \frac{32}{3} L (z) - \frac{112}{9} \right)
  \nonumber \\ 
  & & + \, \frac{64}{3}{\cal D}_2 (z) - \frac{160}{9}{\cal D}_1 (z) + 
  \frac{224}{27}{\cal D}_0 (z)  - \frac{64}{3} L^2 (z) + \frac{448}{9} L (z) - \frac{656}{27} 
\Bigg] \Bigg\} \, ,
\eeqa
which again agrees with an exact calculation, including non-logarithmic NLP terms. 
It is not in fact surprising that this happens: one may derive the next-to-soft Feynman
rules by systematically expanding the exact unintegrated amplitude in the emitted gluon 
momenta (following the diagrammatic approach of Ref.~\cite{Laenen:2010uz}). Thus, 
the effective approach and the full calculation agree at the amplitude level by construction.
We have also checked that, upon combining our results for double real emission and 
real-virtual corrections with the well-known two-loop virtual corrections, and with mass
factorisation, the complete result of Refs.~\cite{Hamberg:1991np,Matsuura:1991pc} is
reproduced, to the expected accuracy.

%%%%%%%%%%%%%%%%%%%%%%%%%%%%%%%%%%%%%%%%%%%%%

\section{Conclusion}
\label{sec:conclude}

In this paper, we have derived an all-order factorisation formula, \eq{NEfactor_nonabel}, 
organising at the amplitude level all contributions which give rise to threshold logarithms
up to next-to leading power. The formula has been derived for the Drell-Yan process, but 
we expect it to apply, with minor modifications, for all processes involving the annihilation
of QCD partons into electroweak final states, such as (multiple) Higgs production via gluon
fusion or multiple vector boson production. \eq{NEfactor_nonabel} generalises the well-known
leading power soft-collinear factorisation formula described in Ref.~\cite{Dixon:2008gr}, 
as well as previous formulae that included only abelian-like contributions~\cite{DelDuca:1990gz,
Bonocore:2015esa}. It contains similar universal functions, namely the leading-power soft 
and jet functions, together with a radiative jet function. We have generalised the definition 
of the latter to a non-abelian theory, and calculated this quantity at one-loop order for quark 
jets.

We have verified our formula by reproducing known threshold logarithms at NNLO in Drell-Yan 
production, which is a non-trivial check at loop-level since both collinear and soft momentum
regions are tested. We discussed how to remove the double counting of next-to-soft and 
collinear contributions via a subtraction term, by defining a {\it next-to-soft radiative jet
function}. We note that a more general definition of this function, in particular for general 
values of the auxiliary vector $n$, deserves further study, which we postpone to future work. 
As was the case for the abelian-like contributions in previous work, we find that there is a 
non-vanishing loop-level contribution to NLP logarithms from hard collinear configurations 
of virtual gluons: this leads to a breaking of the next-to-soft theorems discussed for example 
in Refs.~\cite{Cachazo:2014fwa,Casali:2014xpa}, at loop level.

A new feature of the present work is the role of next-to-soft web diagrams, describing the 
correlated emission of gluons external to the hard interaction. Using next-to-soft webs, we 
reproduce the double real emission contributions in Drell-Yan production, which are 
dominated near threshold by radiation which is always (next-to-)soft, whether or not it is 
collinear. We note that, as discussed in detail in Refs.~\cite{Laenen:2008gt,Laenen:2010uz}, 
the web language is potentially much more powerful, implying formal exponentiation of 
next-to-soft contributions in a much more general context, as discussed for example in 
Refs.~\cite{Gatheral:1983cz,Frenkel:1984pz,Sterman:1981jc,Mitov:2010rp,Gardi:2010rn,
Gardi:2011yz,Gardi:2013ita,Gardi:2011wa,Falcioni:2014pka,Gardi:2013saa,Almelid:2015jia}. 
We conclude that \eq{NEfactor_nonabel} is an important step towards a general resummation 
procedure for NLP threshold logarithms. Further necessary ingredients include the calculation 
of radiative jet functions for external gluon jets, and the elucidation of subleading collinear 
effects in processes with final state parton jets (see, for example~\cite{Nandan:2016ohb}). 
These developments will be the subject of future work.

%%%%%%%%%%%%%%%%%%%%%%%%%%%%%%%%%%%%%%%%%%%%%

\section*{Acknowledgments}

This research was supported by the Research Executive Agency (REA) of
the European Union through the contract PITN-GA-2012-316704
(HIGGSTOOLS), and by by MIUR (Italy), under contract
2010YJ2NYW$\_$006. LV is supported by the People Programme (Marie
Curie Actions) of the European Union Horizon 2020 Framework
H2020-MSCA-IF-2014, under REA grant N.~656463 (Soft Gluons).
DB and EL have been supported by the Netherlands
Foundation for Fundamental Research of Matter (FOM) programme 156,
``Higgs as Probe and Portal'', and by the National Organisation for
Scientific Research (NWO).  This research was also supported in part
by the National Science Foundation under Grant No. NSF PHY-1125915,
and EL is grateful to the Kavli Institute for Theoretical Physics,
where part of this work was done, for hospitality.  CDW is supported
by the UK Science and Technology Facilities Council (STFC).  We are
very grateful to the Higgs Centre for Theoretical Physics at the
University of Edinburgh, where part of this work was carried out, for
warm hospitality. We also thank Simon Caron-Huot, Einan Gardi, and
Jort Sinninghe-Damst\'{e} for useful discussions.

%%%%%%%%%%%%%%%%%%%%%%%%%%%%%%%%%%%%%%%%%%%%%

\bibliography{refs.bib}

%%%%%%%%%%%%%%%%%%%%%%%%%%%%%%%%%%%%%%%%%%%%%

\end{document}